\documentstyle{mn}
\begin{document}

\title[Nonlinear r-modes]{Nonlinear r-modes in a spherical shell:
issues of principle}

\author[Yu. Levin \& G. Ushomirsky]
	{Yuri Levin$^{(1),(2)}$ and Greg Ushomirsky$^{(2)}$ \\
	$^{(1)}$Theoretical Astrophysics, California Institute of
	Technology, Pasadena, California 91125 \\
	$^{(2)}$ Department of Astronomy, 601 Campbell Hall, 
	University of California, Berkeley, California 94720}

\maketitle

\begin{abstract}
We use a simple physical model to study the nonlinear behaviour of
the r-mode instability. We assume that r-modes (Rossby waves) are
excited in a thin spherical shell of rotating incompressible fluid.
For this case, exact Rossby wave solutions of arbitrary amplitude are
known. We find that:

(a) These nonlinear Rossby waves carry ZERO physical angular momentum
and positive physical energy, which is contrary to the folklore belief
that the r-mode angular momentum and energy are negative. We think
that the origin of the confusion lies in the difference between
physical and canonical quantities.

(b) Within our model, we confirm the differential drift reported by
Rezzolla, Lamb and Shapiro~(1999). \newline
Radiation reaction is introduced into the model by assuming that the
fluid is electrically charged; r-modes are coupled to electromagnetic
radiation through current (magnetic) multipole moments.  We study the 
coupled equations of charged fluid and Maxwell field dynamics and find
that:

(c) To linear order in the mode amplitude, r-modes are subject to the
CFS instability, as expected.

(d) Radiation reaction decreases the angular velocity of the shell and
causes differential rotation (which is distinct from but similar in
magnitude to the differential drift reported by Rezzolla et al.) prior
to saturation of the r-mode growth. This is contrary to the
phenomenological treatments to date, which assumed that, prior to the
saturation of the r-mode amplitude, the loss of stellar angular
momentum is accounted for by the r-mode growth.  This establishes, for
the first time, that radiation reaction leads not only to overall loss
of angular momentum, but also to differential rotation.

(e) We show that for $l=2$ r-mode
 electromagnetic radiation reaction is equivalent to
gravitational radiation reaction in the lowest post-Newtonian
order. Based on our electromagnetic calculations, we conclude that
inertial frame dragging, both from the background rotation and from
the r-mode itself, will modify the r-mode frequency by a factor $\sim
R_{\rm Schwarzschild}/R_{\rm star}$, in qualitative agreement with
Kojima~(1998).
\end{abstract}

\section{introduction}

Andersson (1998) has shown, and Friedman and Morsink (1998) have
confirmed analytically, that r-modes of rotating stars can grow
because of gravitational radiation reaction. Lindblom et al.\ (1998)
have shown that this instability can be important in rapidly rotating
hot neutron stars, where the r-mode amplitude might become large
enough to affect the spin frequency [see also Andersson et al.\
(1998)].  The details of nonlinear evolution, which allows for large
r-mode amplitudes, are essential for astrophysical applications [Owen
et al.\ (1998), Spruit(1999), Levin (1999)]; yet only phenomenological
treatments of such nonlinear evolution exist so far. This paper is an
attempt to learn about the nonlinear behaviour of rotating fluid in
which r-modes are driven by radiation reaction.

To address issues of principle, we choose to study a very simple
system. Our ``star'' is a thin rotating shell of incompressible
inviscid fluid which is sandwiched between two hard spheres.  These
spheres exert no friction on the fluid; their role is to make sure
that the fluid motion is restricted to a two-dimensional spherical
surface.

For such thin rotating shell, exact Rossby wave solutions%
\footnote{From here onward ``Rossby waves'' and ``r-modes'' will be used
interchangeably.}
of the fully nonlinear fluid equations are known from the geophysical
literature [see, e.g., Silberman (1954)].  In Section II we review
these solutions and study their properties. We find that in our model
Rossby waves carry zero physical angular momentum and positive
energy. This is somewhat surprising, since Friedman and Morsink (1998)
have shown that CANONICAL angular momentum and energy of the r-modes
are negative. It seems that one cannot equate canonical and physical
quantities for r-modes; one can only do it if the Lagrangian
perturbation of vorticity is zero [Friedman and Schutz, 1978(a)]. This
is not the case for r-modes; this fact is intricately connected to the
differential drift found by Rezzolla et al.\ (1999) and confirmed
within our model. Our results do not argue against the presence of the
CFS instability, since its derivation as given in Friedman and Schutz
[1978(b)] relies only on canonical quantities.

In Section III we switch on radiation reaction by assuming that the
fluid is electrically charged.  We derive coupled dynamical fluid ---
Maxwell field equations [see Eqs (\ref{maineq1})---(\ref{chi2})].  By
studying the linear part of these equations, we then explicitly show
the presence of the CFS instability for r-modes. We also show that the
r-mode frequency is modified by radiation reaction; the frequency
shift is given by $\Delta\omega\sim \gamma_{\rm CFS}(\lambda/R_{\rm
star})^{2l+1}$, where $l$ is the r-mode multipole order, $\gamma_{\rm
CFS}$ is the r-mode growth rate and $\lambda$ is the wavelength of the
emitted gravitational wave [cf Eq.\ (\ref{ratio})].

We study the evolution of the star once the nonlinear
coupling terms are included in the dynamical equations. We show that,
to second order in the r-mode amplitude, radiation reaction slows down
the star and causes it to rotate differentially, and that this is not
related to the saturation of the r-mode. This is contrary to the
phenomenological model introduced by Owen et al.\ (1998), and used
extensively by Levin (1999). In that model it was assumed that, prior
to saturation and in absence of viscosity, the loss of angular
momentum to gravitational radiation is entirely accounted for by the
r-mode growth.  However, since (in our model) r-modes do not carry
physical angular momentum, our finding only seems logical.  The
differential rotation is similar in magnitude, but different in origin
from the differential drift found by Rezzolla et al.\ (1999). Both the
slow-down and the differential rotation are described by Eq.\
(\ref{psi2}) of the text. 

Section IV provides a quick and somewhat superficial excursion into
gravitational radiation reaction potential for mass current quadrupole
moment, in
the $3.5$ post-Newtonian order. We use the formalism of Blanchet
(1997) and others to show that in this order there is a full
equivalence between the gravitational and electromagnetic radiation
reaction. All of our results are thus expected to apply to the case of
Rossby waves coupled to gravitational radiation, at least for
$l=2$. In particular, gravitational radiation reaction
slows down the star and causes it to rotate differentially,
prior to saturation of the r-mode growth. 
By
carrying through results of Section III, we find that inertial frame
dragging both from the background rotation and from the r-mode itself
modifies the r-mode frequency by a factor of $\sim R_{\rm
Schwarzschild}/R_{\rm star}$, in qualitative agreement with results of
Kojima (1998).

\section{Rossby waves in a rotating shell: exact solutions}

Thin rotating shells, and Rossby waves in them, have been studied
extensively by geophysicists and meteorologists since the end of last
century.  In this section we follow closely the work by Silberman
(1954); more original references can be found in that work.

Consider a thin spherical sheet of incompressible fluid of radius $a$,
sandwiched between two spherical hard covers, so that motion of the
fluid is restricted to a two-dimensional spherical surface. The fluid
is rotating with angular frequency $\Omega$ around the $z$-axis
relative to an inertial observer. Since the fluid motion is restricted
to two dimensions, and the fluid is assumed to be incompressible, the
fluid velocity field in the co-rotating frame is completely determined
by a stream function $\psi(\theta,\phi)$ defined on the fluid sphere:
\begin{equation}
v_{\phi}={1\over a}{\partial\psi\over\partial\theta}\hbox{  and  }
v_{\theta}=-{1\over a\sin\theta}{\partial\psi\over\partial\phi},
\label{velocity}
\end{equation}
where $v_{\phi}$ and $v_{\theta}$ are the components of the fluid
velocity along parallels and meridians, respectively.  The vorticity
of the fluid in the rotating frame of reference (called {\it relative}
vorticity) is given by
\begin{equation}
\eta=(1/a)^2\nabla_a^2\psi,
\label{defvorticity}
\end{equation}
where $\nabla_a^2$ is the Laplacian operator on a unit sphere:
\begin{equation}
\nabla_a^2={1\over \sin{\theta}}\left[{\partial\over\partial\theta}\left(
\sin
\theta
{\partial\over\partial\theta}\right)+{1\over \sin{\theta}}{\partial^2\over
\partial
\phi^2}\right].
\label{nabla}
\end{equation}

For inviscid, barotropic%
\footnote{The exact condition for circulation conservation
is $\nabla\rho\times\nabla p=0$, where
$\rho$ and $p$ are the fluid density and pressure respectively.  A
fluid with a one-parameter equation of state, $p=p(\rho)$, satisfies
this condition automatically.}
flows, Euler's fluid equations imply conservation of circulation:
\begin{equation}
\int_{C(t)}\vec{v}^{\rm in}\cdot d\vec{r}=\hbox{const},
\label{ciculation}
\end{equation}
where $C(t)$ is a contour moving with the fluid and $\vec{v}^{\rm in}$
is the velocity relative to an inertial observer.  We derive the
dynamical equations for an in incompressible spherical fluid shell
from this equation.  In particular, Eq.~(\ref{ciculation}) implies
that the radial component of the {\it absolute} vorticity is
conserved:
\begin{equation}
{d\over dt}(\eta+2\Omega\cos\theta)=0
\label{vorteqn1}
\end{equation}
[see, e.g., Longuet-Higgins (1964)]. Here $d/dt=\partial/\partial
t+\vec{v}\cdot\nabla$ is the Lagrangian time derivative.  We would
like to stress that Eq.~(\ref{vorteqn1}) is fully nonlinear, and the
only assumptions made so far are those of incompressibility and zero
viscosity.

By expressing all velocity components and $\eta$ in terms of the
stream function [cf Eqs.~(\ref{velocity}) and~(\ref{defvorticity})],
one can write Eq.~(\ref{vorteqn1}) as a dynamical nonlinear equation
for the stream function evolution:
\begin{equation}
{\partial\over\partial t}
 \nabla_a^2\tilde{\psi}+2\Omega{\partial\over\partial\phi}\tilde{\psi}
 ={1\over \sin{\theta}}\{\nabla_a^2\tilde{\psi},\tilde{\psi}\},
\label{maineqn}
\end{equation}
where $\tilde{\psi}=\psi/a^2$ is the reduced stream function, and the
brackets are defined by $\{A,
B\}=\partial_{\theta}A\partial_{\phi}B-\partial_{\theta}B
\partial_{\phi}A$.  The left-hand side of Eq.\ (\ref{maineqn})
contains only linear terms, while the right-hand side is the nonlinear
advection term.

Let us neglect this nonlinear part for a moment and look for a
solution to the linearized equation in the form
\begin{equation}
\tilde{\psi}=\alpha\Omega Y_{lm}e^{i\omega t}+cc.,
\label{rmode}
\end{equation}
where $\alpha$ is a constant. It is then easy to work out the dispersion
relation for the linear part of Eq.\ (\ref{maineqn}):
\begin{equation}
\omega={2m\Omega\over l(l+1)}.
\label{dispersion}
\end{equation}
This is the well-known dispersion relation for Rossby waves.
Moreover, since $Y_{lm}$ is an eigenfunction of the Laplacian, the
nonlinear term in Eq.\ (\ref{maineqn}) equals zero when $\psi\propto
Y_{lm}$. Therefore Eq.\ (\ref{rmode}) is an {\it exact} solution of
the fluid equations of motion, for arbitrarily large $\alpha$.

Consider now the angular momentum and energy of the solution
(\ref{rmode}).  The total angular momentum of the fluid shell
 is given by
\begin{equation}
L_{\rm star}=L_{\rm background}+L_{\rm r-mode},
\label{lstar}
\end{equation}
where $L_{\rm background}$ is the  angular momentum
of the shell when
the r-mode amplitude $\alpha$ is zero,
and 
\begin{equation}
L_{\rm r-mode}=\rho a^3 \int v_{\phi}\sin^2\theta d\theta d\phi 
\label{lmode}
\end{equation}
is the physical change in angular momentum of the shell due to the
r-mode.  Here $\rho$ is the surface density of the fluid. It is then
clear from Eqs (\ref{velocity}) and (\ref{rmode}) that
\begin{equation}
L_{\rm r-mode}=0
\label{amomentum}
\end{equation}
for all $m\neq0$. Modes with $m=0$ have zero frequency and are called
{\it zonal currents}; we will come back to them later when we discuss
nonlinear evolution of the r-modes.  On the other hand, the canonical
angular momentum of a Rossby wave with $l=m$, computed from Eq.\ (3.4)
of Owen et al.\ (1998), is
\begin{equation}\label{l-canonical}
L_{\rm canonical}=-\frac{[l(l+1)]^2}{2}\rho a^4 \alpha^2 \Omega,
\end{equation}
 clearly different from the physical angular momentum.

The total kinetic energy of the shell
\footnote{For an incompressible spherical fluid shell,
both internal and potential energies are constant,
and hence do not play any dynamical role.} is given by
\begin{equation}
E=\rho a^2\int\left[(a\Omega\sin\theta+v_{\phi})^2+v_{\theta}^2
\right]\sin\theta d\theta d\phi.
\label{energy}
\end{equation}
For r-modes with $m$ not equal to zero, the terms linear in
$v$ on the right-hand side of Eq.\ (\ref{energy}) vanish after
integration over $\phi$. Therefore, the total energy of the star
with the r-mode is
\begin{eqnarray}
E &=& E_{\rm background}+E_{\rm r-mode} \\ \nonumber
&=&\rho a^2\int (a\Omega\sin\theta)^2\sin\theta d\theta d\phi \\ \nonumber
&+&  \rho a^2\int\left(v_{\theta}^2 
+ v_{\phi}^2\right)\sin\theta d\theta d\phi.
\label{energy1}
\end{eqnarray}
The physical r-mode energy $E_{\rm r-mode}$ is therefore greater than
zero. Both this and $L_{\rm r-mode}=0$ [see Eq.\ (\ref{amomentum})] is
contrary to the common belief, which holds that $E_{\rm r-mode}$ and
$L_{\rm r-mode}$ are both negative.

The source of misunderstanding is the confusion between canonical and
physical quantities, which are not equal to each other for r-modes.
In their seminal work, Friedman and Schutz (1978a) have shown that
canonical quantities are equal to the physical ones, so long as the
Lagrangian change in vorticity is zero to second order in the
perturbation amplitude. The last condition cannot be true for r-modes;
this fact is intricately connected to the work by Rezzolla, Lamb and
Shapiro (1999). These authors track the motion of a fluid particle for
the case when the fluid stream function is given by $\psi\propto
Y_{22}$. They find that the particle experiences a drift along stellar
latitude; the speed of the drift depends on the latitude. Therefore,
the authors argue, the presence of an r-mode implies the presence of a
differential drift in the star%
\footnote{In private communications, Rezzolla, Lamb, and Shapiro had
found some of relativity community to be sceptical about the reality
of their claimed differential drift. This scepticism was due to the
fact that Rezzolla, Lamb, and Shapiro derived the differential drift
using the fluid velocities which were first-order quantities in the
r-mode amplitude; however, the drift that they found was second-order
in the r-mode amplitude. The sceptics thought that such procedure was
inconsistent. However, for the case of a spherical shell, the fluid
velocities of the exact Rossby wave solution are STRICTLY linear with
respect to the r-mode amplitude, and are given by exactly the same
expressions as used by Rezzolla, Lamb, and Shapiro. Therefore, at
least for the spherical shell, the differential drift due to the
r-mode is real.}.
This is clearly incompatible with the zero Lagrangian vorticity
perturbations; therefore, the Friedman-Schutz condition is not
satisfied and one cannot equate canonical and physical energy and
angular momentum%
\footnote{One other obvious example where there is no equivalence
between physical and canonical quantities is the shear sound wave in a
solid elastic body. The physical motion of all the particles of the
body is transverse to the direction of the wave propagation; therefore
such wave has zero linear momentum along the direction of
propagation. Yet, the canonical linear momentum is not zero (in fact,
it is $p=N\hbar k$, where $N$ is the number of phonons in the wave,
and $k$ is the wave vector).}.

\section{Electromagnetic radiation reaction for r-modes}

Friedman and Schutz (1978b) had shown that the CFS instability occurs
whenever there is a mode with negative canonical angular momentum,
which is dragged forward by the stellar rotation relative to an
inertial observer, and which is coupled to some radiation field. This
radiation field can be scalar, electromagnetic or gravitational---no
matter what its nature, the CFS instability will be
present. Papaloizou and Pringle (1978) investigated the CFS
instability for f-modes coupled to a scalar field, and derived the
growth rate of an unstable mode by introducing explicitly scalar-field
radiation reaction.

In this section we introduce electromagnetic radiation reaction (ERR)
into our model.  In the next section we will show that, for r-modes on
a spherical shell, gravitational radiation reaction (in the lowest
appropriate post-Newtonian order) and ERR are equivalent.  Thus, all
results derived in this section for the case of the ERR are applicable
to the case of gravitational radiation reaction.

We assume that the fluid is homogeneously electrically charged, with
the charge $q$ per unit mass%
\footnote{Note that this situation is different from the usual MHD,
where the current is due to relative motion of oppositely charged
species, e.g., electrons and protons. Here, the charge current density
is proportional to the mass current density.}. 
R-modes couple to electromagnetic radiation through time-varying 
current multipole
moments (since the fluid is assumed to be incompressible, the charge
multipole moments are zero).  We now derive dynamical equations of
motion of the fluid shell coupled to an electromagnetic field.

The circulation 
around a closed contour moving with the fluid is no longer conserved:
\begin{equation}
{d\over dt}\int_{C(t)} \vec{v}^{\rm in}\cdot d\vec{r}=
\int_{C(t)}{d\vec{v}^{\rm in}\over dt}\cdot d\vec{r}+
\int_{C(t)} \vec{v}^{\rm in}\cdot d\left({d\vec{r}\over dt}\right).
\label{circ1}
\end{equation}
Here, as in Eq.\ (\ref{ciculation}), $\vec{v}^{\rm in}$ is the fluid
velocity in the inertial frame of reference.  The second term on the
RHS of Eq.\ (\ref{circ1}) is identically zero; the first one is in
general not zero because of forces exerted on the fluid by the Maxwell
field:
\begin{eqnarray}
{d\over dt}\int_{C(t)} \vec{v}^{\rm in}\cdot d\vec{r} &=&
q\int_{C(t)} \vec{E}\cdot d\vec{r}+{q\over c}\int_{C(t)} 
\vec{v}^{\rm in}\times\vec{B}\cdot
d\vec{r} \nonumber \\
&=& -{q\over c}{d\over dt}\int_{C(t)} \vec{B}\cdot \vec{dA}.
\label{circ2}
\end{eqnarray}
Here $\int_{C(t)} \vec{B}\cdot \vec{dA}$ is the magnetic flux through
the surface with boundary $C$, with $d\vec{A}$ being the area
increment vector. Equation (\ref{circ2}) is the well-known Faraday's
law, which states that the electromotive force (EMF) around a closed
contour equals to the negative of the rate of change of the magnetic
flux through the contour.  We see that the presence of a Maxwell field
modifies Eq.\ (\ref{ciculation}) so that
\begin{equation}
\hbox{circulation}+(q/c)\cdot\hbox{flux}=\hbox{const}.
\end{equation}
The analogue of Eq.\ (\ref{vorteqn1}) is then
\begin{equation}
{d\over dt}(\eta+2\Omega\cos\theta+{q\over c}B_r)=0,
\label{vorteqn2}
\end{equation}
where $B_r$ is the radial component of the magnetic field in the
inertial frame.  Expressing velocity components in terms of the stream
function derivatives [cf. Eq.\ (\ref{velocity})], we get
\begin{eqnarray}
{\partial\over\partial t}\nabla_a^2\tilde{\psi}+
2\Omega{\partial\over \partial\phi}\tilde{\psi}+{q\over c}
{\partial B_r\over 
\partial t}
&=& {1\over \sin\theta}\left\{\nabla_a^2\tilde{\psi},
\tilde{ \psi}\right\} \nonumber \\ &+&
{q\over c\sin\theta}\left\{B_r, \tilde{\psi}\right\}.
\label{maineq1}
\end{eqnarray}
The angular operator $\nabla_a^2$ is given by Eq.\ (\ref{nabla}).
Equation (\ref{maineq1}) governs the fluid motion under the action of
an electromagnetic field, which, in turn, is generated by the fluid
currents.

Let us consider a Rossby wave with a reduced stream function given by
$\tilde{\psi}_{lm}=\tilde{\psi}^+_{lm}+\tilde{\psi}^-_{lm}$, where
\begin{equation}
\tilde{\psi}^+_{lm}=\left(\tilde{\psi}^-_{lm}\right)^*=\alpha\Omega
Y_{lm}e^{i\omega_{lm} t}.
\label{psilm}
\end{equation}
Then, as shown in Appendix A [cf. Eqs (\ref{Bro1}), (\ref{Brot1}),
(\ref{blm1}), (\ref{chi11}), and (\ref{chi21})], the radial component
$B_r$ of the magnetic field at the fluid shell is
\begin{equation}
B_r=B_{\rm rot}+B_{lm},
\label{Bro}
\end{equation}
where
\begin{equation}
B_{\rm rot}={8\pi q\over 3c}
\rho\Omega a\cos\theta
\label{Brot}
\end{equation}
is the radial component of a dipole magnetic field due to the uniform
rotation of the charged fluid shell, and
\begin{equation}
B_{lm}=\chi\tilde{\psi}^+_{lm}+\chi^*\tilde{\psi}^-_{lm}
\label{blm}
\end{equation}
is the radial component of the magnetic field produced by the Rossby
wave itself. Here $\chi=\chi_1+i\chi_2$, where, to lowest order in
$ka$, we have
\begin{eqnarray}
\chi_1&=&-{4\pi l(l+1)\over 2l+1}{q\rho a\over c},\label{chi1}\\
\chi_2&=&{4\pi l(l+1)\over [(2l+1)!!]^2}{q\rho a\over c}(ka)^{2l+1}.
\label{chi2}
\end{eqnarray}
The wavevector $k$ of the emitted electromagnetic waves is given by
\begin{equation}
k={1\over c}(\omega_{lm}-m\Omega) 
=-{m\over l}{(l-1)(l+2)\over l+1}{\Omega\over c}.
\label{k}
\end{equation}
We now obtain the new dispersion relation for Rossby waves by
substituting Eq.\ (\ref{Bro}) into Eq.\ (\ref{maineq1}) and keeping
terms which are linear in the mode amplitude $\alpha$. The resulting
linear equation is
\begin{equation}
{\partial\over\partial t}\nabla_a^2\tilde{\psi}_{lm}+
2\Omega{\partial\over \partial\phi}\tilde{\psi}_{lm}+
{q\over c\sin\theta}\left\{\tilde{\psi}_{lm}, B_{\rm rot}\right\}+
{q\over c}{\partial B_{lm}\over \partial t}=0.
\label{maineq3}
\end{equation} 
Using Eqs (\ref{Brot}), (\ref{blm}), (\ref{chi1}) and (\ref{chi2}),
we derive the Rossby-wave dispersion relation [cf. Eq.~(\ref{dispersion})]:
\begin{equation}
\omega_{lm}={2\Omega m\over l(l+1)}{1+(1/3)\epsilon\over
1+\epsilon/(2l+1)-i\epsilon (ka)^{2l+1}/[(2l+1)!!]^2},
\label{omegalm1}
\end{equation}
where
\begin{equation}
\epsilon={4\pi q^2\rho a\over c^2}={Q^2/a\over Mc^2}.
\label{epsilon}
\end{equation}
Here $Q$ and $M$ are the charge and the mass of the fluid shell
respectively; $\epsilon$ is thus the ratio of the energy of
electrostatic self-interaction and the rest-mass energy of the
shell. For the case when a Rossby wave is coupled to gravitational
radiation, we will see that the analogue of $\epsilon$ is $\sim R_{\rm
Schwarzschild}/R_{\rm star}$ in the weak gravity regime. For a neutron
star, $\epsilon\sim0.4$. Then, to first order in $\epsilon$, the
angular frequency of the r-mode is
\begin{equation}
\omega_{lm}={2\Omega m\over l(l+1)}\left[1+{\Delta\omega_{lm}\over\omega_{lm}}
-i{\gamma_{\rm CFS}\over\omega_{lm}}\right],
\label{omegalm2}
\end{equation}
where
\begin{equation}
\Delta\omega_{lm}={2(l-1)\over 3(2l+1)}\epsilon\omega_{lm}
\label{deltaomega1}
\end{equation}
is the frequency shift of the r-mode due to the electromagnetic
interaction, and
\begin{equation}
\gamma_{\rm CFS}=-{\epsilon (ka)^{2l+1}\over [(2l+1)!!]^2}\omega_{lm}
\label{gamma1}
\end{equation}
is the growth rate of the r-mode due to the CFS instability.  The
angular frequency shift and the CFS growth rate are related by
\begin{equation}
{\gamma_{\rm CFS}\over\Delta\omega_{lm}}\sim\left({\Omega a\over c}
\right)^{2l+1}\sim \left({a\over \lambda}\right)^{2l+1},
\label{ratio}
\end{equation} 
where $\lambda$ is the wavelength of the emitted radiation.

Now we consider the terms which are of second order with respect to
the r-mode amplitude $\alpha$ in the dynamical
equation~(\ref{maineq1}). The advection term (first term on RHS of
Eq.~[\ref{maineq1}]) does not contribute to this order, but the second
term on the RHS of the Eq.~(\ref{maineq1}) does have a component which
is proportional to $\alpha^2$:
\begin{eqnarray}
{\rm nonlinear\ term}
&=&{q\over c\sin\theta}\left\{B_{lm},\tilde{\psi}_{lm}\right\}  \\
&=& {q\over c\sin\theta}\left\{\chi\tilde{\psi}^+_{lm}
	+\chi^*\tilde{\psi}^-_{lm},
\tilde{\psi}^+_{lm}+\tilde{\psi}^-_{lm}\right\}\nonumber\\ \nonumber
&=&
2i\alpha^2\Omega^2 Im(\chi) \frac{q}{c\sin\theta}
\left\{Y_{lm},Y^*_{lm}\right\}.
\label{nonlinearterm}
\end{eqnarray}
Let us focus, for concreteness, on the $l=m=2$ mode, which has the
largest growth rate (and hence is perhaps the most important
astrophysically). Then the nonlinear term in Eq.~(\ref{nonlinearterm})
becomes
\begin{equation}
{\rm nonlinear\ term}={18}\alpha^2\Omega\gamma_{\rm CFS}
\left(\sqrt{9\over 7\pi}Y_{30}-\sqrt{3\over \pi}Y_{10}\right).
\label{nonlinearterm1}
\end{equation}
This term on the right-hand side of Eq.\ (\ref{maineq1}) will act as a
source for the left-hand side of this equation, thus creating a
contribution to the stream function which is second order in $\alpha$:
\begin{equation}
\tilde{\psi}_{(2)}=\kappa_1(t)Y_{10}+\kappa_2(t)Y_{30},
\label{secondorder}
\end{equation}
where, to leading order in $\epsilon$, $\kappa_1$ and $\kappa_2$
satisfy the following evolution equations:
\begin{eqnarray}
{d\kappa_1\over dt}
&=&
{9 }\sqrt{3\over \pi}\alpha^2\Omega\gamma_{\rm CFS},
\nonumber\\
{d\kappa_2\over dt}
&=&
-{9\over 2}\sqrt{1\over 7\pi}\alpha^2\Omega\gamma_{\rm CFS}.
\label{kappa}
\end{eqnarray}
If the mode amplitude grows exponentially starting from a small value
$\alpha_0$, i.e., $\alpha=\alpha_0 e^{\gamma_{\rm CFS}t}$, then by
integrating Eq.\ (\ref{kappa}) we get the following expression for the
reduced stream function of the radiation reaction induced flow:
\begin{equation}
\tilde{\psi}_{(2)}
\simeq {9\over 2}\sqrt{3\over \pi}\alpha^2\Omega Y_{10}-
{9\over 4}\sqrt{1\over 7\pi}\alpha^2\Omega Y_{30}.
\label{psi2}
\end{equation}
Let us discuss this equation. The first term on the right-hand side
represents a uniform flow in the direction opposite to stellar
rotation.  {\it It thus represents the spindown of the star.}  The
angular momentum associated with this flow is
\begin{equation}
L_{\rm spindown}=-18\rho a^4\alpha^2\Omega,
\end{equation}
in agreement with the canonical angular momentum of the r-mode itself
[cf. Eq.~(\ref{l-canonical})]. This is to be expected, as both $L_{\rm
canonical}$ and $L_{\rm spindown}$ must be equal to the angular
momentum lost to radiation.

The second term in Eq.~(\ref{psi2}) is a zonal current or, in another
words, differential rotation.  This term does not contribute to the
angular momentum of the star. This differential rotation is similar in
magnitude to the differential drift reported by Rezzolla et
al. (1999). However, its origin is completely different.  In our case,
the differential rotation is driven by the radiation reaction, whereas
the differential drift of Rezzolla et al. is a kinematic property of
the r-mode fluid motion, and is not at all related to radiation
reaction.

Note that both the uniform slow-down and the induced differential
rotation are not related directly to the r-mode saturation, and are
present well before saturation takes place. This is somewhat contrary
to the phenomenological model of nonlinear behaviour of the r-mode
instability by Owen et al. (1998), which was used extensively by Levin
(1999).  This model assumed that prior to r-mode saturation, the loss
of angular momentum and energy carried off by gravitational waves was
manifested by the r-mode growth, and that the background motion of the
star was unchanged. The intuition for this model was based heavily on
the belief that r-modes carry negative {\it physical} energy and
angular momentum.  We now know that the latter is in general not true,
so it should not be surprising that radiation reaction induces
second-order changes in the background motion of the star, {\it as
well as} drives the r-mode instability.  Astrophysical implications of
this point are currently under investigation, and will be the subject
of our next publication.

\section{Gravitational radiation reaction for $l=2$ r-modes}
An analogy between weak gravity and electromagnetism
has been studied by many researchers
[e.g., Braginsky, Caves and Thorne (1977),
Thorne, Price and Macdonald (1986)].
Shapiro (1996) has shown that the Newtonian
circulation around a closed contour comoving
with a perfect
fluid
is not conserved in presence of a 
gravitomagnetic field; the conserved quantity
(termed ``relativistic circulation'' by Shapiro)
is a linear combination of the Newtonian
circulation around the contour and a gravitomagnetic flux
through the contour [cf.\ Eq.\ (4) of Shapiro (1996)].
This is very reminiscent of our conclusions 
for the circulation of a charged fluid in the presence
of magnetic field; in fact, the two derivations are
almost identical. Since it is the circulation equation
that determines the dynamics of our rotating shell,
Shapiro's result is of great relevance to understanding this
dynamics.

In this section we use results of Blanchet (1997), Shapiro (1996),
Asada et al.\
(1997), and Rezzolla et al.\ (1998) to investigate the effect of
gravitational radiation reaction on r-modes in a spherical fluid
shell. We find that, for slow-motion systems, there is a close analogy
with electromagnetic radiation reaction considered in the previous
section.

Following Asada et al.\ (1997) and Rezzolla et al.\ (1998), we consider
a $3+1$ splitting of spacetime.  We write the square of the line element
as
\begin{eqnarray}
ds^2 &=& g_{\mu\nu}dx^{\mu}dx^{\nu} \\ \nonumber
&=&(\alpha^2-\beta_i\beta^i) c^2 dt^2+2\beta_i cdt dx^i+\gamma_{ij}dx^i dx^j,
\label{metric}
\end{eqnarray}
where $\alpha$ and $\beta^i$ are the lapse function and the shift
vector respectively, and $\gamma_{ij}$ are the spatial metric
coefficients.  For weakly gravitating ($R\gg R_{\rm Schwarzschild} $),
slow-motion ($v\ll c$) sources in which mass currents produce
gravitational radiation, one can choose a gauge such that it is the
time-varying shift vector $\vec{\beta}$ that plays a dynamically
important role, relative to all other perturbations of the metric.
For a periodic mass-current quadrupole moment, the shift vector
consists of two parts, $\vec{\beta}=\vec{\beta^{\rm
gm}}+\vec{\beta^{\rm rr}}$, where \newline
1. The first part is the usual gravitomagnetic vector; its leading
term in $v/c$ is given by
\begin{equation}
\vec{\beta^{\rm gm}}(\vec{r})=-{4G\over c^3}
\int\sigma {\vec{v}^{\rm in}(\vec{r^{\prime}})
\over |\vec{r}-\vec{r^{\prime}}|} d^3 r^{\prime}
,
\label{bgm}
\end{equation}
where $\vec{v}^{\rm in}$ is the fluid speed relative to an inertial
observer, and $\sigma$ is the mass volume density of the source [cf.\ Eq.\
(3.4) of Blanchet (1997) when $c=\infty$; Blanchet uses a vector
potential $\vec{V}=- \vec{\beta}/4$].
\newline
2. The second part is responsible for the radiation reaction; it
changes sign under time reversal. Its leading term in $v/c$ is given
by [cf.\ Eq.\ (3.66) of Blanchet (1997) and Eq.\ (17.6) of Rezzolla et
al.\ (1998)]:
\begin{equation}
\beta_i={16G\over 45c^8}\epsilon_{ijk}x_ix_j S^{(5)}_{kl},
\label{betai}
\end{equation}
where
\begin{equation}
S_{ij}=\int d^3x\epsilon_{kl(i}x_{j)} x_k \sigma v_l^{\rm in}
\label{sij}
\end{equation}
is the mass-current quadrupole moment. Here the superscript $(n)$
stands for $d^n/dt^n$, and the brackets around tensorial indices,
$(ij)$, indicate symmetrization over these indices
(i.e. $a_{(ij)}=1/2(a_{ij}+a_{ji})$ for a tensor $a_{ij}$).

The effective force $\vec{F}$ per unit mass that this shift metric
perturbation exerts on the fluid is given by
\begin{equation}
\sigma^{-1}\vec{F}=-{c}{\partial\vec{\beta}\over\partial t}+
{c}\vec{v^{\rm in}}\times\nabla\times\vec{\beta},
\label{star}
\end{equation}
cf.\ Eq.\ (12) of Rezzolla et al.\ (1998) and Eq.\ (1) of Shapiro (1996).
  Note that this expression
for the gravitational force is equivalent the Lorentz force exerted by
an electromagnetic field on moving charged fluid:
\begin{equation}
\sigma^{-1}\vec{F}_{\rm em}=-{q\over c}{\partial\vec{A}\over\partial t}+
{q\over c}\vec{v^{\rm in}}\times\nabla\times\vec{A}.
\label{star1}
\end{equation}
In Equations (\ref{star}) and (\ref{star1}), the shift vector $\beta$
is dynamically equivalent to $c^2 q\vec{A}$, where $\vec{A}$ is the
electrodynamical vector potential.

In our discussion of the motion of charged fluid on a spherical shell,
we have shown that it was the radial component of the magnetic field
that entered the dynamical equations of the fluid motion. Likewise, an
identical argument will work for the gravitational force given by Eq.\
(\ref{star}). Therefore, the radial component of
$\nabla\times\vec{\beta}$ enters the equations of motion of a
gravitating fluid:
\begin{equation}
\partial_t\nabla_a\tilde{\psi}+2\Omega{\partial\tilde{\psi}\over
\partial\phi}+{c}{\partial b\over \partial t}=
{1\over \sin\theta}\left\{\nabla_a\tilde{\psi}, \tilde{\psi}\right\}+
{c\over {\sin\theta}}\left\{b, \tilde{\psi}\right\},
\label{star3}
\end{equation}
where $b=(\nabla\times \vec{\beta})_r$.

Suppose that 
 a single $l=2$ Rossby wave is excited in a spherical shell with
the surface mass density $\rho$, and that the wave's
reduced stream function is given by $\tilde{\psi}_{2m}=\alpha\Omega Y_{2m}e^{i\omega_{2m}
t}$. Then, as is shown in Appendix B, in the slow-motion approximation
the radial component of the gravitomagnetic
field generated by the fluid motion 
is given by
\begin{equation}
b=b_{\rm rot}+b_{2m};
\label{gravmag}
\end{equation}
cf. Eq.\ (\ref{Bro}). Here 
\begin{equation}
b_{\rm rot}=-{4\over 3}\epsilon_{\rm grav}{\Omega\over c}\cos\theta,
\label{bgmrot}
\end{equation}
and
\begin{equation}
b_{2m}=\chi_{\rm grav}\tilde{\psi}_{2m},
\label{blmgrav}
\end{equation}
where, to lowest order in $ka$,     
\begin{eqnarray}
Re\left(\chi_{\rm grav}\right)&=&{12\epsilon_{\rm grav}\over 3c},
\label{chi1grav}\\
Im\left(\chi_{\rm grav}\right)&=&{48\epsilon_{\rm grav}\over 225 c}
(ka)^5,\label{chigrav2}
\end{eqnarray}
and
\begin{equation}
\epsilon_{\rm grav}={2GM\over c^2 a}={R_{\rm Schwarzschild}\over a}.
\end{equation}
Equations (\ref{gravmag}), (\ref{bgmrot}), and (\ref{blmgrav}) have
the same structure as the analogous equations for the electromagnetic
case, cf. Eqs (\ref{Bro}), (\ref{Brot}), and (\ref{blm}). By following
the same steps as in the electromagnetic case, we work out the
dispersion relation for $l=2$ Rossby waves interacting with gravity:
\begin{equation}
\omega_{2m}=
{2\Omega\over 3}\left(1+{\Delta\omega_{2m}\over \omega_{2m}}-i{\gamma_{\rm CFS}\over \omega_{2m}}
\right),
\label{omegalmgrav}
\end{equation}
where
\begin{equation}
\Delta\omega_{2m}\simeq -{4\over 15}\epsilon_{\rm grav}\omega_{2m},
\label{deltaomegagrav}
\end{equation}
is the shift of the r-mode frequency due to inertial frame dragging, which
originates both from stellar rotation and from the mode itself; and
\begin{equation}
\gamma_{\rm CFS}\simeq -\omega_{2m}{cIm\left(\chi_{\rm grav}\right)\over 6}=
\omega_{2m}{8\epsilon_{\rm grav}|ka|^5\over 225 }
\label{gammacfsgrav}
\end{equation}
is the growth rate of the r-mode due to the CFS instability driven by
the gravitational radiation reaction. This growth rate agrees with the
calculations of Lindblom et al.\ (1998) when one applies their Eq.\
(17) to the case of a massive spherical shell.

The r-mode frequency shift due to inertial frame dragging was
discovered by Kojima (1998). For the case of a real three-dimensional
star, Kojima (1998) claims, and Beyer and Kokkotas (1999) confirm,
that such shift causes the r-mode spectrum to be continuous.  This
claim, however, is not supported by calculations of Lockitch,
Andersson and Friedman (1999). The issue of a continuous spectrum is
not relevant for the spherical shell.

Since the formalisms for a slow-motion gravitational radiation
reaction from a mass-current quadrupole and for an electromagnetic
radiation reaction from a charge-current quadrupole are identical in
the structure of the dynamical equations, all of the conclusions from
the previous section about the nonlinear electromagnetically-driven
evolution of the $l=2$ r-mode are also valid for a gravitationally
driven $l=2$ r-mode, at least for the case when the r-modes are
excited in a spherical shell.  More specifically, Eq.\ (\ref{psi2})
for the secondary reaction-induced flow is still valid. Therefore,
gravitational radiation reaction will slow down the star and cause it
to rotate differentially; the  former will account for the
loss of angular momentum to gravitational waves.  Both the slow-down
of the star and the reaction-induced differential rotation are not
related to the nonlinear saturation of the r-mode growth.

\section{conclusions and bets}

This paper has studied the issues of principle for the r-mode
instability in the nonlinear regime. Although we considered only a
special case of r-modes excited in a spherical rotating shell, we bet
that most of the lessons learned from the simple model will apply to
real stars. In particular, r-modes in general do not carry negative
{\it physical} angular momentum and energy; the radiation reaction
causes the star to slow down and rotate differentially prior to the
r-mode saturation. Therefore, phenomenological nonlinear evolution
Equations (3.14)~---~(3.17) of Owen et al.\ (1998) need to be
reconsidered.

We believe thus that our conclusions argue in favor of Spruit's
conjecture (1999) that the r-mode instability causes differential
rotation in the star.  Spruit has modeled differential rotation as
relative motion of two spherical shells, while we find differential
rotation in the lateral direction. We believe that, in the
three-dimensional case, both radial and lateral differential rotation
will develop.  A more detailed discussion of this and other
astrophysical consequences of our formalism is a subject of a future
publication.

We want to thank Luciano Rezzolla, Frederick Lamb, and Stuart Shapiro
for showing us the advance draft of their manuscript, and for detailed
comments on our first draft. Discussions with John Friedman, Peter
Goldreich, Keith Lockitch, and Kip Thorne have been very helpful.
Alan Wiseman has pointed us to useful references on radiation
reaction.  YL was supported by the NSF grant AST-9731698, and by the
Theoretical Astrophysics Center at UCB; GU is a Fannie and John Hertz
Foundation Fellow.

\appendix

\section{Magnetic field generated by
a Rossby wave in a charged fluid.}

In this Appendix we find the radial component of the magnetic field
generated by a Rossby wave in a charged fluid. We thus derive Eqs
(\ref{Bro})---(\ref{chi2}) of the text.

Suppose that a single Rossby wave is excited in a rotating shell, and
that its reduced stream function is given by
\begin{equation}
\tilde{\psi}_{lm}=\tilde{\psi}^+_{lm}+\tilde{\psi}^-_{lm}
=\alpha\Omega Y_{lm}
e^{i\omega_{lm}t}
+\alpha^*\Omega Y_{lm}^*e^{-i\omega_{lm}t}.
\label{psi10}
\end{equation}
The radial component of the magnetic field produced by such fluid
motion can be found by using the multipole formalism discussed in
Sec.\ 16.5 of Jackson (1975). In particular, using Eq.\ (16.87) of
this reference, we find:
\begin{equation}
B_r=B_{\rm rot}+B_{lm},
\label{Bro1}
\end{equation}
where
\begin{equation}
B_{\rm rot}={q\rho\over c}
\int {2\Omega\cos\theta\over |\vec{r}-
\vec{r^{\prime}}|}a^2\sin\theta d\theta d\phi
,
\label{brot1}
\end{equation}
and 
\begin{equation}
B_{lm}={q\rho\over ac}\int {e^{-ik|\vec{r}-
\vec{r^{\prime}}|}\over |\vec{r}-
\vec{r^{\prime}}|}\vec{r^{\prime}}\cdot\nabla
\times \vec{v} a^2\sin\theta^{\prime} d\theta^{\prime} d
\phi^{\prime}.
\label{blm11}
\end{equation}
Here $B_{\rm rot}$ is the dipole magnetic field due to the uniform
rotation of the charged shell, while $B_{lm}$ is the field due to the
Rossby wave.  The wavenumber $k$ is that of the emitted
electromagnetic radiation:
\begin{equation}
k={1\over c}(\omega_{lm}-m\Omega).
\label{k1}
\end{equation}
Equations (\ref{brot1}) and (\ref{blm11}) are evaluated by noting that
$\vec{r}\cdot\nabla\times\vec{v}= a\nabla_a^2\tilde{\psi}_{lm} =
-al(l+1)\tilde{\psi}^+_{lm}+c.c.$, and that
\begin{eqnarray}
\int{e^{-ik|\vec{r}-
\vec{r^{\prime}}|}\over |\vec{r}-
\vec{r^{\prime}}|}Y_{lm}(\theta^{\prime}, \phi^{\prime})
\sin\theta^{\prime} d\theta^{\prime}d\phi^{\prime} \\ \nonumber =
4\pi ik h^{(1)}_l(ka)j_l(ka)Y_{lm}(\theta, \phi)
,
\label{spherharm}
\end{eqnarray}
where $j_l$ and $h^{(1)}_l$ are spherical Bessel and Hankel functions
respectively (see e.g. Eqs (16.9) and (16.10) of Jackson (1975)). After
some algebra, we get
\begin{equation}
B_{\rm rot}={8\pi q\over 3c}
\rho\Omega a\cos\theta,
\label{Brot1}
\end{equation}
and
\begin{equation}
B_{lm}=\chi\tilde{\psi}^+_{lm}+\chi^*\tilde{\psi}^-_{lm}
\label{blm1}
\end{equation}
 Here $\chi=\chi_1+i\chi_2$, where, to lowest order in $ka$,
\begin{eqnarray}
\chi_1&=&-{4\pi l(l+1)\over 2l+1}{q\rho a\over c},\label{chi11}\\
\chi_2&=&{4\pi l(l+1)\over [(2l+1)!!]^2}{q\rho a\over c}(ka)^{2l+1}.
\label{chi21}
\end{eqnarray} 
Thus, we have derived Equations (\ref{Bro}), (\ref{Brot}),
(\ref{blm}), (\ref{chi1}), and (\ref{chi2}) of the text.

\section{Gravitomagnetic field generated by an $l=2$ Rossby
wave in a thin shell}

In this Appendix we derive Eqs (\ref{gravmag}), (\ref{blmgrav}),
(\ref{chi1grav}), and (\ref{chigrav2}) for the radial component of the
gravitomagnetic field, $b=(\nabla\times\vec{\beta})_r$, generated by a
rotating massive shell in which an $l=2$ Rossby wave is excited.

Evaluation of the part of $b$ which is not responsible for radiation
reaction, to leading order in $v/c$, is straightforward.  The relevant
part of the shift vector, $\vec{\beta}_{\rm nonradiative}$, is given
by Eq.\ (\ref{bgm}) of the text, which, up to a constant factor, is
same as the nonradiative part of the electromagnetic vector potential:
\begin{equation}
\vec{A}_{\rm nonradiative}={1\over c}\int {\sigma q \vec{v}^{\rm in}
(r^{\prime})\over |r-r^{\prime}|} d^3r^{\prime}.
\label{Anr}
\end{equation}
Therefore,
\begin{eqnarray}
\vec{\beta}_{\rm nonradiative}&=&-{4G\over c^2 q}\vec{A}_{\rm nonradiative},
\nonumber\\
b_{\rm nonradiative}&=&-{4G\over c^2 q}{B_r}_{\rm nonradiative}.\label{nonrad}
\end{eqnarray}
Here, as in the text, $\sigma$ is the mass {\it volume} density of the
fluid (trivial changes must be made for the case of a two-dimensional
sphere), and $q$ is the charge per unit mass of the fluid.  By using
Eqs. (\ref{Bro}), (\ref{Brot}), (\ref{blm}), and (\ref{chi1}) of the
text, and substituting $l=2$, we get
\begin{equation}
b_{\rm nonradiative}=-{4\over 3}\epsilon_{\rm grav}{\Omega\over c}\cos
\theta+{12\epsilon_{\rm grav}\over 3c}\tilde{\psi}_{2m},
\label{bnr}
\end{equation}
where $\epsilon_{\rm grav}=2GM/(c^2 a)=R_{\rm Schwarzschild}/a$.

Now we shall derive the expression for the part of the radial
gravitomagnetic field which is responsible for radiation reaction; we
shall denote it by $b^{\rm rr}$.  The radiative part of the shift
vector is given by Eq.\ (\ref{betai}) of the text, which can be
rewritten as follows:
\begin{eqnarray}
{\beta}_i^{\rm rr}&=&\left[{8G\over 45 c^7}
\epsilon_{ijk}x_jx_l
\int\sigma\left( J_k^{\prime}x_l^{\prime}+J_l^{\prime}
x_k^{\prime}\right)d^3 x^{\prime}\right]^{(5)
},
\end{eqnarray}
or, in index-free form,
\begin{equation}
\vec{\beta}^{\rm rr}=
\left\{{8G\over 45 c^7}
\int\sigma\left[\left(\vec{r}\cdot
\vec{r}^{\prime}\right)\left(\vec{r}\times
\vec{J}^{\prime}\right)
+\left(\vec{r}\cdot\vec{J}^{\prime}\right)
\left(\vec{r}\times\vec{r}^{\prime}\right)\right]
d^3x^{\prime}\right\}^{(5)},
\label{betarad}
\end{equation}
where
\begin{equation}
\vec{J}^{\prime}=\vec{r}^{\prime}\times\vec{v}
(\vec{r}^{\prime})=-\nabla{\psi}(\vec{r}^{\prime})
.
\label{J}
\end{equation}
In this expression, vectors $\vec{r}$, $\vec{r}^{\prime}$, and the
fifth time derivative are defined relative to an inertial observer at
rest. The radial component of the radiation reaction gravitomagnetic
field is given by
\begin{equation}
b^{\rm rr}=
{1\over a}(\vec{r}\cdot
\nabla\times\vec{\beta}^{\rm rr})=
\left[
{48G\sigma\over 45 c^8 a}
\int\sigma (\vec{r}\cdot
\vec{r}^{\prime})[\vec{r}\cdot\nabla\psi(\vec{r}
^{\prime})]d^3 x^{\prime}\right]^{(5)},
\label{b}
\end{equation}
or, for the case of a spherical shell,
\begin{equation}
b={48G\rho a\over 45 c^8}\left[\int
(\vec{r}\cdot
\vec{r}^{\prime})[\vec{r}\cdot\nabla\psi(\vec{r}
^{\prime})]\sin\theta^{\prime} d\theta^{\prime}
d\phi^{\prime}\right]^{(5)}.
\label{b1}
\end{equation}

Let the stream function $\psi$ be a linear combination of $l=2$
spherical harmonics:
\begin{equation}
\psi(\theta, \phi)=\sum_{m=-2}^2 a_{2m}Y_{2m}(\theta, \phi).
\end{equation}
Let us choose a coordinate system $x_1, y_1, z_1$ (with polar angles
$\theta_1$ and $\phi_1$), such that the axis $z_1$ is along $\vec{r}$.
In this system, $\psi$ is also a linear of $l=2$ spherical harmonics,
but with different weight coefficients $a_{2m}^{1}$:
\begin{equation}
\psi(\vec{r}_1)=\sum_{m=-2}^2 a_{2m}^{1}Y_{2m}(\theta_1, \phi_1).
\end{equation}
For $\vec{r}_1=\vec{r}$ (i.e.\ for $\theta_1=0$),
\begin{equation}
\psi(\vec{r})=a_{20}^1Y_{20}(0, \phi_1)=\sqrt{5\over 4\pi}a_{20}^1.
\label{psinew}
\end{equation}
In the new coordinate system we can now evaluate $b(\vec{r})$:
\begin{eqnarray}
b(\vec{r})&=&{48\pi G\rho a^3\over 45 c^7 } \\ \nonumber &\times&
\left\{\int \cos\theta_1\left[
\vec{r}\cdot\nabla \sum_{m=-2}^2 a^1_{2m}Y_{2m}(\theta_1,
\phi_1)\right]\sin\theta_1 d\theta_1 d\phi_1\right\}^{(5)}.
\label{b11}
\end{eqnarray}
In Eq.\ (\ref{b11}), all terms with non-zero $m$ vanish after
integration over $\phi$; only the term with $m=0$ contributes to the
integral.  Integrating the remaining term, we get
\begin{equation}
b(\vec{r})=\left[{48G\rho\over 45 c^8}{8\over 5}\pi a^4 \sqrt{5\over 4\pi}a_{20}^1
\right]^{(5)}.
\label{b12}
\end{equation}
Substituting Eq.\ (\ref{psinew}) into Eq.\ (\ref{b12}),
and evaluating the fifth time derivative,
we get
\begin{equation}
b(\vec{r})=i{48\over 225}\epsilon_{\rm grav}(ka)^5{\tilde{\psi}_{2m}\over c}.
\label{brr}
\end{equation}
Here $\epsilon_{\rm grav}=2GM/(c^2a)=R_{\rm Schwarzschild}/a$.
Equations (\ref{bnr}) and (\ref{brr}) are equivalent to Equations
(\ref{gravmag}), (\ref{bgmrot}), (\ref{blmgrav}), (\ref{chi1grav}),
and (\ref{chigrav2}) of the text.
 

\begin{thebibliography}{}
\bibitem{bib1}
Andersson, N.\ 1998, ApJ, 502, 708.
\bibitem{bib2}
Andersson, N., Kokkotas, K.\ D.\ \& Schutz, B.\ F.\
1999, ApJ, 516, 307.
\bibitem{bib3}
Asada, H., Shibata, M., \& Futamase, T., Prog.\
Theor.\ Phys.\ Suppl., 96, 81.
\bibitem{bib4}
Beyer, H.\ R., Kokkotas, K.\ D.\ 1999, MNRAS 308, 745.

\bibitem{bib5}
Blanchet, L.\  1997, Phys.\ Rev.\ D, 55, 714.
\bibitem{bib6}
Braginsky, V.\ B., Caves, C.\ M., \& Thorne, K.\ S.\
1977, Phys.\ Rev.\ D, 2047.

\bibitem{bib7}
Friedman, J.\ F., \& Morsink, S.\ M.\ 1998, ApJ, 502, 714.
\bibitem{bib8}
Friedman, J.\ F., \& Schutz, B.\ F.\ 1978a, ApJ, 221, 937.
\bibitem{bib9}
Friedman, J.\ F., \& Schutz, B.\ F.\ 1978b, ApJ, 222, 281.
\bibitem{bib10}
Kojima, Y.\ 1998, MNRAS, 293, 423.
\bibitem{bib11}
Levin, Y.\ 1999, ApJ, 517, 328.
\bibitem{bib12}
Lindblom, L., Owen, B.\ J., \& Morsink, S.\ M.\ 1998,
Phys.\ Rev.\ Lett., 80, 4843.
\bibitem{bib13}
Lockitch, K.\ H.,  Andersson, N., \& Friedman, J.\ F.\ 1999,
unpublished work; preliminary results are reported in
K.\ H.\ Lockitch's PhD thesis, gr-qc/9909029.
\bibitem{bib14}
Longuet-Higgins, M.\ S.\ 1964, Proc.\ Roy.\ Soc.\ London A, 279, 446.

\bibitem{bib15}
Owen, B.~J., Lindblom, L., Cutler, C., Schutz, B.~F., Vecchio, A., \&
  Andersson, N.\ 1998, Phys.\ Rev.\ D, 58, 084020

\bibitem{bib16}
Papaloizou, J., \& Pringle, J.\ E.\ 1978, MNRAS, 184, 501.

\bibitem{bib17}
Rezzolla, L., Lamb, F.\ K., \& Shapiro, S.\ L.\ 1999,
submitted to ApJ, astro-ph/9911188.

\bibitem{bib18}
Rezzolla, L., Shibata, M., Asada, H., Baumgarte, T.W., \& Shapiro, S.\
to appear in ApJ (gr-qc/9905057).

\bibitem{bib19}
Shapiro, S.\ L.\ 1996, Phys.\ Rev.\ Lett., 77, 4487.
\bibitem{bib20}
Silberman, I.\ 1954, Journal of Meteorology, 11, 27.

\bibitem{bib21}
Spruit, H.\ C.\ 1999, A\&A, 341, L1.
\bibitem{bib22}
Thorne, K.\ S., Price, R.\ H., \& Macdonald,
D.\ A.\ 1986, {\it Black Holes: the Membrane
Paradigm} (Yale University Press, New Haven),
Chapter 4.

\end{thebibliography}
\end{document}